\documentclass
[letterpaper,english,pra,aps,preprint,nofootinbib,superscriptaddress]{revtex4}%
\usepackage{graphicx}
\usepackage{color}
\usepackage{bm}
\usepackage{amsmath}
\usepackage{amssymb}
\usepackage{amsfonts}%

\begin{document}

\title{The Quantum Cheshire Cat effect: Theoretical basis and observational implications}
\author{Q.\ Duprey}
\affiliation{Laboratoire de Physique Th\'{e}orique et
Mod\'{e}lisation (CNRS Unit\'{e} 8089), Universit\'{e} de
Cergy-Pontoise, 95302 Cergy-Pontoise cedex, France}
\author{S. Kanjilal}
\affiliation{Center for Astroparticle Physics and Space Science
(CAPSS), Bose Institute, Kolkata 700 091, India}
\author{U. Sinha}
\affiliation{Raman Research Institute, Sadashivanagar, Bangalore,
India}
\author{D. Home}
\affiliation{Center for Astroparticle Physics and Space Science
(CAPSS), Bose Institute, Kolkata 700 091, India}
\author{A.Matzkin}
 \affiliation{Laboratoire de Physique Th\'{e}orique et
Mod\'{e}lisation (CNRS Unit\'{e} 8089), Universit\'{e} de
Cergy-Pontoise, 95302 Cergy-Pontoise cedex, France}

\begin{abstract}
The Quantum Cheshire Cat 
(QCC) is an effect introduced recently within the Weak Measurements framework.
The main feature of the QCC effect is that a property of a quantum particle appears to be spatially
separated from its position. The status of this effect has however
remained unclear, as claims of experimental observation of the QCC
have been disputed by strong criticism of the experimental as well
as the theoretical aspects of the effect. In this paper we clarify
in what precise sense the QCC can be regarded as an unambiguous
consequence of the standard quantum mechanical formalism applied to
describe quantum pointers weakly coupled to a system. In light of
this clarification, the raised criticisms of the QCC effect are
rebutted. We further point out that the limitations of the
experiments performed to date imply that a loophole-free
experimental demonstration of the QCC has not yet been achieved.

\end{abstract}
\maketitle

\section{Introduction}

Weak measurements were introduced \cite{AAV} in 1988 as a theoretical scheme
for minimally perturbing non-destructive quantum measurements. It is indeed
well-known that a standard quantum measurement irremediably destroys the
premeasurement state of a system by projecting it to an eigenstate of the
measured observable, say $B$. It is therefore impossible, according to this
``standard view'', to know anything about a given system property, represented
by an observable $A$, as the system evolves from a prepared initial state
towards the final eigenstate obtained after measuring $B$.

The weak measurement scheme introduced by Aharonov, Albert and Vaidman \cite{AAV} aims to measure $A$
without appreciably modifying the system evolution relative to the case without the measurement. This is achieved by means of a weak coupling between the system
observable $A$ and a dynamical variable of an external degree of freedom (that we will designate here by the term ``quantum pointer''). This weak coupling entangles the
quantum pointer with the system, until the final projective measurement of $B$
correlates the obtained system eigenstate with the quantum state of the weak
pointer. The resulting state of the weak pointer has picked up a shift
(relative to its initial pre-coupled state) proportional to the real part of a quantity known
as the \emph{weak value } of $A$.

While the meaning of weak values has been debated since their
inception \cite{leggett89,peresWM,repAAV,soko2016}, several experimental
implementations of the weak measurement protocol have been carried
out: weak values have thus been measured for different observables
in many quantum systems \cite{photonWM,sterling2009,
lundeen,torres2014,A2015PRA,delayest,cormann2017,wvest}. Concurrently,
theoretical schemes based on weak measurements have been proposed
with practical \cite{soko2013,wmtrap,romito2016} or
foundational
\cite{A2012PRL,pati2014,aharonov-ap15,hoffmann2016,QDAM17} aims.
Among the latter, Aharonov et al \cite{aharonov-qcc} introduced an
interferometric based scheme baptized the Quantum Cheshire Cat
(QCC). A QCC situation takes place when at some location (say,
region $I$) the weak value of an observable representing a system
property (eg, polarization) vanishes while the weak value of the
system's spatial projector is non-zero.\ At some other location (say
region $II$) the opposite takes place (the weak values of the
spatial projector and system property are zero and non-zero respectively).
Echoing the features of Lewis Carroll's eponym character, Aharonov
et al. loosely described this Cheshire cat situation in Ref.\
\cite{aharonov-qcc} as seeing the grin (the polarization) without
the cat (the photon in region $I$)\footnote{As we will see below,
the property (the grin) is not the polarization itself but a
specific polarization component. A generalization that would apply
to any polarization component was suggested in Ref.
\cite{completeQCC}.}; they further wrote in \cite{aharonov-qcc} that
the QCC scheme implies ``\emph{physical properties can be
disembodied from the objects they belong to}''. A tentative
experimental realisation of the QCC effect employing neutrons
published soon after \cite{A2014NatComm} concluded ``\emph{that the
system behaves as if the neutrons go through one beam path, while
their magnetic moment travels along the other}''. Very recently, an
experiment with single photons similar to the neutron one was
reported \cite{QCC-photon}.

The Quantum Cheshire cat phenomenon reported in Refs
\cite{aharonov-qcc,A2014NatComm} has raised a string of criticisms,
in particular in the published works
\cite{CCNJP,deraedt2015,atherton2015,stuckey2016,soko2016}. The common trend
in many of these criticisms is to view the QCC effect as a false paradox, an
illusion that would have an arguably simpler interpretation.
Unfortunately, rather than making their point using unambiguous
conceptual and technical terms, most of the works criticizing the Quantum
Cheshire Cat effect did not analyze the weak measurements framework
(generally even avoiding to mention weak values). Interpretational claims (that ultimately depend on the
interpretation of the standard quantum formalism, not on the
specificities of weak values) were not always distinguished from practical pointer readouts, and
the theoretical QCC effect was not carefully discriminated from the
tentative experimental implementations.

Hence, instead of demystifying the Quantum Cheshire Cat, the criticism that
has been made may have brought additional confusion. To be fair, it must be pointed out that some of
the well-known works dealing with weak measurements, such as the QCC paper
\cite{aharonov-qcc} put the stress on delivering a simple take-home message
without providing the detailed formal arguments that would explain and justify
the main message.

In this work, our aim will be to carefully scrutinize the Quantum Cheshire Cat
effect in order to lift the confusion on what this effect is really about. We
will discriminate the theoretical, ideal QCC from the shortcomings that can
inevitably appear in any experimental realization. We will disentangle the
interpretational aspects from the technical terms in which the effect is
couched. We will in particular argue through the analysis of the meaning of
vanishing weak values that spatial separation between a particle and its
properties can be consistently defined, provided one is willing to relax
ascribing properties to a quantum system solely through the
eigenstate-eigenvalue link.

In order to do so we will first recall the aspects of weak measurements that
will be relevant to describe the Quantum Cheshire Cat (Sec.\ 2).\ We will then
define, in Sec.\ 3, the QCC effect from the weak values obtained when precise
coupling conditions are met, detailing the weak measurement process that was
omitted in the original QCC paper \cite{aharonov-qcc} or in subsequent QCC
related works \cite{A2013JPA}.\ Indeed, a proper account of this detailed
process appears to be one of the two crucial ingredients necessary in order to
dispel the confusion that has emerged around the QCC effect. Sec.\ 4 will be
devoted to the description of the tentative experimental implementation of the
QCC with neutrons and with single photons.\ We will in particular highlight
several essential differences with the ideal theoretical scheme. The published
criticism of the QCC effect will be examined in Sec.\ 5 and compared to the
precisely defined QCC introduced in Sec.\ 3. In the Discussion section
(Sec.\ 6) we will detail the technical and conceptual aspects of the arguments
given in the criticisms; we will also explain under which assumptions it is
legitimate to interpret the effect as \textquotedblleft
disembodiment\textquotedblright.\ This will be seen to depend on the general
interpretation of the quantum formalism that is favoured. We will finally give
our conclusions in Sec.\ 7; anticipating on our assessments, we will conclude
that: (i) the Quantum Cheshire Cat effect is a well defined quantum feature
derived from the standard quantum formalism that can be interpreted as a spatial separation of a particle from one of its properties if some assumptions regarding property ascription are made; (ii) the QCC effect as predicted
theoretically has not yet been experimentally observed; (iii) most of the works
rebuking the QCC effect produced a substantial criticism of the experimental
attempts to observe the Quantum Cheshire Cat, but a criticism of the ideal
QCC can only be undertaken within a proper conceptual framework able to account for the
relation between weakly coupled pointers and the properties of the measured quantum system. This
is in our view necessary in order to analyze the issue of spatial separation
of a quantum particle from one of its properties in a pre and postselected situation.

\section{Weak measurements and weak values\label{wv-f}}

\subsection{Protocol: Preselection, Unitary coupling, Postselection}

The basic idea underlying the weak measurement (WM) approach is to
give an answer to the question:\textquotedblleft\emph{what is the value of a
property of a quantum system at some intermediate time while the system
evolves from an initial state $\left\vert \psi(t_{i})\right\rangle $ to a
final state $\left\vert \chi(t_{f})\right\rangle $ ?}\textquotedblright, where  $\left\vert \chi(t_{f})\right\rangle $ is the result of a standard projective measurement made at time $t_f$.  As we are only interested here in
applying WM to derive the Quantum Cheshire Cat effect, we will restrict our
discussion to the property corresponding to a bivalued observable $A$, with eigenstates and eigenvalues
denoted by $A\left\vert a_{k}\right\rangle =a_{k}\left\vert a_{k}\right\rangle
,$ $k=1,2$.

Suppose that initially (at $t=t_{i}$) the system is prepared (preselected) into the
state $\left\vert \psi(t_{i})\right\rangle $.\ Let $\left\vert \varphi(t_{i})\right\rangle$ designate the initial state of the quantum pointer. The total
initial quantum state is the product state%
\begin{equation}
\left\vert \Psi(t_{i})\right\rangle =\left\vert \psi(t_{i})\right\rangle
\left\vert \varphi(t_{i})\right\rangle .\label{inis}%
\end{equation}
We assume the pointer is local (its wavefunction has compact support in
configuration space), and that the system and the pointer will interact during
a brief time interval $\tau$ centered around $t=t_{w}$ (physically
corresponding to the time during which the system and the quantum pointer
interact). The interaction between the system and the quantum pointer is given by the Hamiltonian%
\begin{equation}
H_{int}=g(t)AP.\label{Hint}%
\end{equation}
$A$ is the system observable that couples to the momentum $P$ of the pointer. $g(t)$
is a smooth function non-vanishing only in the interval $t_{w}-\tau
/2<t<t_{w}+\tau/2$ and such that $g\equiv\int_{t_{w}-\tau/2}^{t_{w}+\tau
/2}g(t)dt$ appears as the effective coupling constant. Recall that the
coupling (\ref{Hint}) is the usual interaction employed to account for
projective measurements of $A$ (von Neumann's impulsive model): in that case
$g(t)$ is sharply peaked and each $\left\vert
a_{k}\right\rangle $ is correlated with an orthogonal state of the strongly coupled pointer. In a projective measurement the collpase of the pointer 
projects the system state to a random eigenstate $\left\vert a_{k_{0}%
}\right\rangle $. Here instead $g$ will be small, the weakly coupled quantum
pointer does not collapse, and the system will undergo at postselection a
genuine projective measurement that will consequently project the quantum
pointer to a specific final state, as we now detail.

Let $U(t_{w},t_{i})$ be the unitary operator describing the system evolution between
$t_{i}$ and $t_{w}$ (we disregard the self-evolution of the pointer state).
After the interaction $(t>t_{w}+\tau/2)$ the initial uncoupled state
(\ref{inis}) has become entangled \footnote{In Eq. (\ref{corrsl20}) the
interaction appears to take place precisely at $t_{w}$; this ``midpoint rule''
holds provided $\tau$ is small relative to the system evolution timescale (see
Appdx A in the Supp. Mat. of Ref. \cite{A2012PRL}).}:%
\begin{align}
\left\vert \Psi(t)\right\rangle  &  =U(t,t_{w})e^{-igAP}U(t_{w},t_{i}%
)\left\vert \psi(t_{i})\right\rangle \left\vert \varphi(t_{i})\right\rangle
\label{corrsl20}\\
&  =U(t,t_{w})e^{-igAP}\left\vert \psi(t_{w})\right\rangle \left\vert
\varphi(t_{i})\right\rangle \label{corrsl2}\\
&  =U(t,t_{w})\sum_{k=1,2}e^{-iga_{k}P}\left\langle a_{k}\right\vert \left.
\psi(t_{w})\right\rangle \left\vert a_{k}\right\rangle \left\vert
\varphi(t_{i})\right\rangle .\label{corrs}%
\end{align}
 At time
$t_{f}$  the system undergoes a standard projective measurement: a given observable $\hat{B}$ is measured and the system ends up in
one of its eigenstates $\left\vert b_{k}\right\rangle $. We filter the
results of this projective measurement by keeping only a chosen outcome, say $b_{f}$. The system is thus postselected in the corresponding state $\left\vert b_{f}\right\rangle $. We will denote the
postselected state by $\left\vert \chi_{f}(t_{f})\right\rangle \equiv
\left\vert b_{f}\right\rangle $; in most of the paper we will be dealing with
a single postselected state and drop the second label, writing $\left\vert
\chi(t_{f})\right\rangle $ or $\left\vert \chi_{f}\right\rangle $
instead.\ After postselection, the state of the pointer, given by Eq. (\ref{corrs}), becomes

\begin{equation}
\left\vert \varphi(t_{f})\right\rangle =\sum_{k=1,2}\left[  \left\langle
\chi(t_{w})\right\vert \left.  a_{k}\right\rangle \left\langle a_{k}%
\right\vert \left.  \psi(t_{w})\right\rangle \right]  e^{-iga_{k}P}\left\vert
\varphi(t_{i})\right\rangle ,\label{finalps}%
\end{equation}
where we have used $\left\langle \chi(t_{w})\right\vert =\left\langle
\chi(t_{f})\right\vert U(t_{f},t_{w})$. $\varphi(x,t_{f})$ is then given by a
superposition of shifted initial states
\begin{equation}
\varphi(x,t_{f})=\sum_{k=1,2}\left[  \left\langle \chi(t_{w})\right\vert
\left.  a_{k}\right\rangle \left\langle a_{k}\right\vert \left.  \psi
(t_{w})\right\rangle \right]  \varphi(x-ga_{k},t_{i}).\label{finalpspos}%
\end{equation}
This expression is similar to the usual von Neumann projective
measurement: the first step of a projective measurement (the
\emph{premeasurement}, in which each eigenstate $\left\vert a_{k}\right\rangle
$ of the measured observable is correlated with a given state $\varphi
(x-ga_{k})$ of the pointer) is identical here, but in a projective measurement
the second step is a projection to an eigenstate $\left\vert a_{k_{f}%
}\right\rangle $ of A.

Let us now assume the coupling $g$ is sufficiently small so that
$e^{-iga_{k}P}\approx1-iga_{k}P$ holds for each $k$\footnote{The exact
condition for the asymptotic expression to hold implies that the higher order
terms of order $m$ obey $\left\Vert g^{m}P^{m}\left\langle \chi(t_{w}%
)\right\vert A^{m}\left\vert \psi(t_{w})\right\rangle \right\Vert
\ll\left\Vert gP\left\langle \chi(t_{w})\right\vert A\left\vert \psi
(t_{w})\right\rangle \right\Vert <1$ for Eq. (\ref{za1}) and $\left\Vert
gPA^{w}\right\Vert \ll1$ for Eq. (\ref{finwv}). These conditions take precise
forms for specific pointer wavefunctions, in particular when $\varphi
(x,t_{i})$ is Gaussian, see eg Ref. \cite{duck89}.}. Eq. (\ref{finalps})
becomes%
\begin{align}
\left\vert \varphi(t_{f})\right\rangle  &  =\left\langle \chi(t_{w}%
)\right\vert \left.  \psi(t_{w})\right\rangle \left(  1-igP\frac{\left\langle
\chi(t_{w})\right\vert A\left\vert \psi(t_{w})\right\rangle }{\left\langle
\chi(t_{w})\right\vert \left.  \psi(t_{w})\right\rangle }\right)  \left\vert
\varphi(t_{i})\right\rangle \label{za1}\\
&  =\left\langle \chi(t_{w})\right\vert \left.  \psi(t_{w})\right\rangle
\exp\left(  -igA^{w}P\right)  \left\vert \varphi(t_{i})\right\rangle
\label{finwv}%
\end{align}
where%
\begin{equation}
A^{w}=\frac{\left\langle \chi(t_{w})\right\vert A\left\vert \psi
(t_{w})\right\rangle }{\left\langle \chi(t_{w})\right\vert \left.  \psi
(t_{w})\right\rangle }\label{wvt}%
\end{equation}
is the weak value of the observable $A$ given pre and postselected states
$\left\vert \psi\right\rangle $ and $\left\vert \chi\right\rangle $
respectively (we will sometimes employ instead the full notation
$A_{\left\langle \chi\right\vert ,\left\vert \psi\right\rangle }^{w}$ to
specify pre and postselection). For a localized pointer state, expanding to
first order the terms $\varphi(x-ga_{k},t_{i})$ in Eq. (\ref{finalpspos})
leads to Eq. (\ref{finwv}): when $A^{w}$ is real, the overall shift
$\varphi(x-gA^{w},t_{i})$ is readily seen to result from the interference due
to the superposition of the slightly shifted terms $\varphi(x-ga_{k},t_{i})$,
as shown early on in Ref. \cite{duck89}. Note that the shift $gA^{w}$ is very small
so it needs to be evaluated from the statistics over a large number of trials.

Summing up, we see that a weak measurement contains 4 steps: preselection, weak coupling through the
Hamiltonian (\ref{Hint}), postselection and readout of the quantum pointer.

\subsection{Weak values}

\subsubsection{Observable average}

$A^{w}$ is in general a complex quantity. Following Eq. (\ref{finwv}) the real
part of the weak value $A^{w}$ appears as the shift brought to the initial
pointer state $\left\vert \varphi(t_{i})\right\rangle $ by the interaction
with the system via the coupling Hamitonian (\ref{Hint}). The weak values
are generally different from the eigenvalues, but obey a similar relation with
regard to the computation of expectation values. Indeed, the expectation value
of $A$ in state $\left\vert \psi(t_{w})\right\rangle $, given in terms of
eigenvalues by%
\begin{equation}
\left\langle \psi(t_{w})\right\vert A\left\vert \psi(t_{w})\right\rangle
=\sum_{f}\left\vert \left\langle a_{f}\right\vert \left.  \psi(t_{w}%
)\right\rangle \right\vert ^{2}a_{f},\label{sav}%
\end{equation}
can also be written in terms of weak values as%
\begin{align}
\left\langle \psi(t_{w})\right\vert A\left\vert \psi(t_{w})\right\rangle  &
=\left\langle \psi(t_{w})\right\vert U^{\dagger}(t_{f},t_{w})U(t_{f}%
,t_{w})A\left\vert \psi(t_{w})\right\rangle \label{expvwv2}\\
&  =\left\langle \psi(t_{w})\right\vert U^{\dagger}(t_{f},t_{w})\sum
_{f}\left\vert b_{f}\right\rangle \left\langle b_{f}\right\vert U(t_{f}%
,t_{w})A\left\vert \psi(t_{w})\right\rangle \\
&  =\sum_{f}\left\vert \left\langle \chi_{f}(t_{f})\right\vert \left.
\psi(t_{f})\right\rangle \right\vert ^{2}\frac{\left\langle \chi_{f}%
(t_{w})\right\vert A\left\vert \psi(t_{w})\right\rangle }{\left\langle
\chi_{f}(t_{w})\right\vert \left.  \psi(t_{w})\right\rangle }\\
&  =\sum_{f}\left\vert \left\langle \chi_{f}(t_{f})\right\vert \left.
\psi(t_{f})\right\rangle \right\vert ^{2}A_{\left\langle \chi_{f}\right\vert
,\left\vert \psi\right\rangle }^{w}\label{expvwv}%
\end{align}
where we have used $\left\langle \chi_{f}(t_{f})\right\vert =\left\langle
b_{f}\right\vert $ and $\left\langle \chi_{f}(t_{w})\right\vert =\left\langle
\chi_{f}(t_{f})\right\vert U(t_{f},t_{w})$ denotes the postselected state
$\left\langle b_{f}\right\vert $ evolved backward in time up to $t=t_{w}$.
Eq.
(\ref{expvwv}) is expressed in terms of the probabilities of obtaining a
postselected state $\left\vert b_{f}\right\rangle $ (instead of the probability of obtaining an eigenstate), with the rationale that the probabilities of obtaining the postselected states are not modified by the weak coupling.

\subsubsection{Vanishing weak values\label{nullwv}}

Let us start by first looking at the case of null	 \emph{eigenvalues}. In a standard
projective measurement, a vanishing eigenvalue implies that the
state of the measurement pointer is left, untouched, remaining in the initial state: the
coupling has no effect on the pointer. But the system state does change, as it
is projected to the eigenstate associated with the null eigenvalue for the
measured observable.\ For example imagine a beam of atoms incoming on a
beamsplitter, after which the quantum state of each atom can be described by
the superposition $\left\vert u\right\rangle +\left\vert l\right\rangle $ of
atoms traveling along the \emph{u}pper or \emph{l}ower paths; if a measurement
of the projector onto the upper path $\Pi_{u}\equiv\left\vert u\right\rangle
\left\langle u\right\vert $ yields 0, then the quantum state has collapsed to
$\left\vert l\right\rangle $ and indeed one is certain to find the atom on the
lower path. If the atom has some integer spin, then measuring the spin
projection along some direction can yield a null eigenvalue.\ The atom spin
state is projected to the corresponding eigenstate (as can be verified by
making subsequent measurements) corresponding to no spin component along that direction.

For weak measurements, the main difference is that the system's state is not
projected after the weak coupling. Instead, the overall state evolves
according to $e^{-ig{A}P}\left\vert \psi(t_{w})\right\rangle \left\vert
\varphi(t_{i})\right\rangle \approx\left\vert \psi(t_{w})\right\rangle
\left\vert \varphi(t_{i})\right\rangle -iA\left(  g\left\vert \psi
(t_{w})\right\rangle \right)  P\left\vert \varphi(t_{i})\right\rangle .$ The
system state appears as essentially unperturbed except for the small fraction
$g\left\vert \psi(t_{w})\right\rangle $ that couples to the quantum pointer.
$A\left(  g\left\vert \psi(t_{w})\right\rangle \right)  $ is precisely the
post-interaction (or premeasured) part of the system state, corresponding to
the slight change in the system state produced by the weak coupling to the
quantum pointer. The weak value appears as the imprint of this coupling left
on the pointer, conditioned on the final projective
measurement.\ A\emph{\ vanishing weak value} correlates successful postselection
with the quantum pointer having been left unchanged despite the interaction
with the system. The reason is that the quantity $\left\langle
\chi(t_{w})\right\vert A\left\vert \psi(t_{w})\right\rangle $, which is the
numerator in the definition (\ref{wvt}) of $A^{w}$ vanishes. We will adopt here 
Feynman's terminology  \cite{FeynmanHibbs} and designate this quantity by the term ``transition element'' \footnote{In the literature, this term is also called ``transition matrix element'' or ``transition amplitude''}.

Hence when a weak value vanishes the coupling
has no effect (though the pointer is left unchanged) as in the case of vanishing eigenvalues, but the
implication is not relative to eigenvectors but to the transition elements
between $A\left(  g\left\vert \psi(t_{w})\right\rangle \right)  $
and the postselected state $\left\vert \chi(t_{w})\right\rangle $.
Put differently, if the postselected state is obtained, then
whenever $A^{w}=0$ the property represented by $A$ cannot be
detected by the weakly coupled quantum pointer. For example when the
weak value of a projector $\left(  \Pi_{u}\right)  _{\left\langle
\chi\right\vert ,\left\vert \psi\right\rangle }^{w}=0$, this implies
(i) no effective action of the coupling on the quantum pointer (that
is left unchanged) with respect to the postselected state
$|\chi_{f}\rangle$ and (ii) that the state $\left\vert
\chi_{f}\right\rangle $ cannot be reached from the initial state
$\left\vert \psi(t_{i})\right\rangle $ by the part of the system
wavefunction that has interacted with the pointer in the region
where $\Pi_{u}$ was weakly measured \footnote{Note that this
reasonings hold for asymptotically weak couplings, those that do not
affect the postselection probabilities $\left\vert \left\langle
\chi_{f}(t_{f})\right\vert \left. \psi(t_{f})\right\rangle
\right\vert ^{2}$; otherwise terms beyond the linear expansion
(\ref{za1}) would contribute.}. If instead some spin observable
$S_{\gamma}$ is weakly measured in some region and $\left\langle
S_{\gamma}\right\rangle _{\left\langle f\right\vert ,\left\vert
i\right\rangle }^{w}=0$ this implies again that (i) there is no
effective action on the quantum pointer of the coupling between
$S_{\gamma}$ and the pointer variable, and (ii) the final spin state
$\left\vert f\right\rangle $ cannot be obtained by premeasuring
$S_{\gamma}$ on the initial spin state $\left\vert i\right\rangle ,$
i.e. the fraction of the spin state that is perturbed by the
interaction with the pointer and thereby transforms as
$S_{\gamma}\left\vert i\right\rangle $ does not reach $\left\vert
f\right\rangle $. We will discuss in Sec.\ \ref{discuss} how these
statements can be further interpreted. We will for the moment note
that the argument encapsulated by (ii) can be logically restated as:
since the final state is reached when postselection is successful, a
null weak value (i.e., a vanishing transition element) implies
that for this postselected system state, the property represented by
the weakly measured observable cannot have been detected by the
quantum pointer. Hence in some loose sense (to be refined below),
that property \textquotedblleft was not there\textquotedblright, ie
in the region where the system and weak pointer interacted.

\begin{figure}[tb]
\includegraphics[height=8cm]{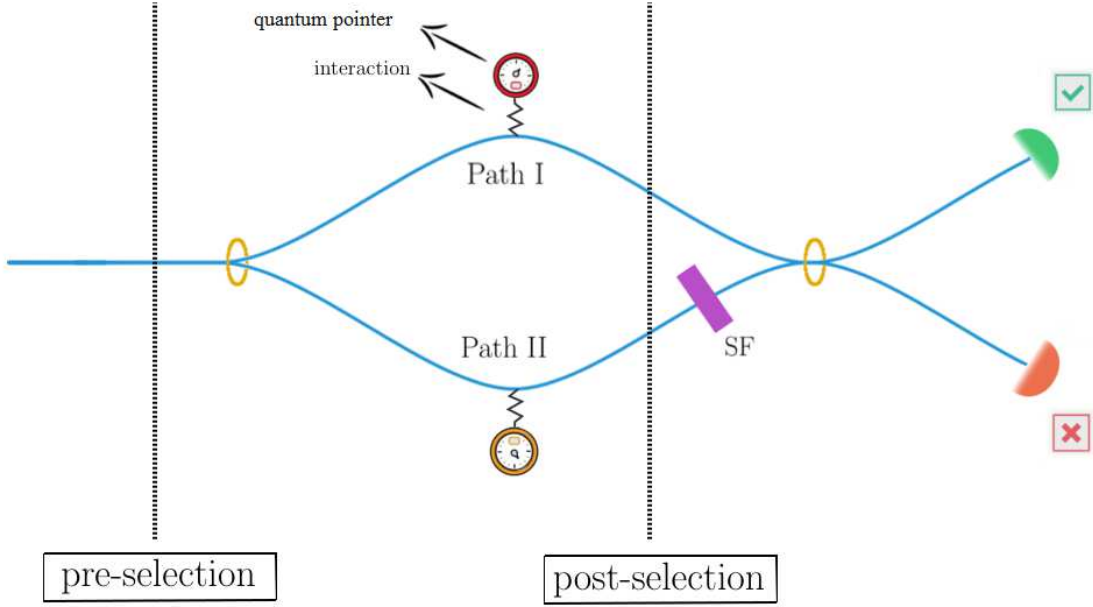}\caption{The ideal Quantum Cheshire Cat
setup based on a 2-path interferometer. A quantum pointer is placed on each
path. Each pointer interacts locally with the particle, coupling the particle
observable $A_{j}=\Pi_{j}$ or $A_{j}=(\sigma_{x})_{j}$ (where $j=I,II$) to the
pointer dynamical variable $P_{I}$ or $P_{II}$ resp. ($g$ is the coupling
strength). The particle then travels along the interferometer essentially
unperturbed until it gets detected (SF is a spin flipper, required for
postselection of the spin state). Successful postselection (detection in the
upper port) is correlated with pointers in final states $\exp\left(
-igA_{j}^{w}P_{j}\right)  \left\vert \varphi_{j}(t_{i})\right\rangle $ with
the weak values $\Pi_{I}^{w}=1,\left(  \sigma_{x}\right)  _{I}^{w}=0$ and
$\Pi_{II}^{w}=0,\left(  \sigma_{x}\right)  _{II}^{w}=1$ for pointers $I$ and
$II$ respectively.}%
\label{fig-MZqcc}%
\end{figure}

\section{The Quantum Cheshire Cat effect\label{qccT}}

The Quantum Cheshire Cat effect, as introduced in Ref. \cite{aharonov-qcc} is
based on the Mach-Zehnder interferometer shown in Fig.\ \ref{fig-MZqcc}.
Rather than elaborating on the original argument \cite{aharonov-qcc}, we will
instead give a proper account of the QCC in terms of weak measurements, by
filling the gaps left in the theoretical account given in \cite{aharonov-qcc}%
.\ We will mostly remain here at a technical level (interpretations will be
discussed later in Sec.\ \ref{discuss}).

Assume that a quantum particle with spin $1/2$ enters the interferometer shown
in Fig.\ \ref{fig-MZqcc} in state $\left\vert \psi_{i}\right\rangle \left\vert
s_{i}\right\rangle $ where $\left\vert \psi_{i}\right\rangle $ is a localized
wavepacket and $\left\vert s_{i}\right\rangle $ the spin state in which the
system was prepared. Let us label by $I$ and $II$ the upper and lower arms of
the interferometer, and assume a quantum pointer in state $\left\vert
\varphi_{I}(t_{i})\right\rangle $ sits on arm $I.$ Hence the total
system-pointer initial state is
\begin{equation}
\left\vert \Psi(t_{i})\right\rangle =\left\vert \psi_{i}\right\rangle
\left\vert s_{i}\right\rangle \left\vert \varphi_{I}(t_{i})\right\rangle
.\label{ini-qcc}%
\end{equation}
The system will then enter the interferometer, evolving to
\begin{equation}
U(t,t_{i})\left\vert \psi_{i}\right\rangle \left\vert s_{i}\right\rangle
=\frac{1}{\sqrt{2}}\left(  \left\vert \psi_{I}\right\rangle +\left\vert
\psi_{II}\right\rangle \right)  \left\vert s_{i}\right\rangle \label{incom}%
\end{equation}
and then locally interact along arm $I$ through the interaction Hamitonian
(\ref{Hint}) given here by $g(t)A_{I}P_{I}$ where index $I$ emphasizes the
coupling takes place only on branch $I$ and $P_{I}$ is the pointer's linear
momentum.\ $A$ is a system observable that will be taken to be either the
spatial projector or some spin component. After the interaction the state
vector is given by Eq. (\ref{corrsl2}) that takes here the form%
\begin{equation}
\left\vert \Psi(t)\right\rangle =U(t,t_{w})\left(  e^{-igA_{I}P_{I}}\left\vert
\psi_{I}\right\rangle \left\vert s_{i}\right\rangle +\left\vert \psi
_{II}\right\rangle \left\vert s_{i}\right\rangle \right)  \left\vert
\varphi_{I}(t_{i})\right\rangle /\sqrt{2}.\label{comp}%
\end{equation}
Assume finally that the system is postselected after exiting the
interferometer ($\left\vert \psi_{I}\right\rangle $ and $\left\vert \psi
_{II}\right\rangle $ thus overlap) in the state%
\begin{equation}
\left\vert \chi_{f}\right\rangle =\left(  \left\vert \psi_{I}\right\rangle
\left\vert s_{i}\right\rangle +\left\vert \psi_{II}\right\rangle \left\vert
-s_{i}\right\rangle \right)  /\sqrt{2}.\label{qcc-ps}%
\end{equation}
To be definite we will choose the initial state to be prepared in state
$\left\vert s_{i}\right\rangle =\left\vert +z\right\rangle $ so $\left\vert
-s_{i}\right\rangle =\left\vert -z\right\rangle .$

Let us now examine the final state of the quantum pointer, that will depend on
the observables to which it was weakly coupled. Then, after postselection, the
pointer state is given by Eq. (\ref{finwv}):%
\begin{align}
\left\vert \varphi_{I}(t_{f})\right\rangle  &  =\frac{1}{\sqrt{2}}\left\langle
\chi_{f}\right\vert U(t_{f},t_{w})\left(  \left\vert \psi_{I}\right\rangle
+\left\vert \psi_{II}\right\rangle \right)  \left\vert s_{i}\right\rangle
\exp\left(  -igA_{I}^{w}P_{I}\right)  \left\vert \varphi_{I}(t_{i}%
)\right\rangle \\
&  =\frac{1}{2}\exp\left(  -igA_{I}^{w}P_{I}\right)  \left\vert \varphi
_{I}(t_{i})\right\rangle .
\end{align}
Hence the final state of the quantum pointer that was weakly coupled to
$A_{I}$ depends on the weak value%
\begin{align}
A_{I}^{w}  &  =\frac{\left\langle \chi_{f}\right\vert U(t_{f},t_{w}%
)A_{I}U(t_{w},t_{i})\left\vert \psi_{i}\right\rangle \left\vert s_{i}%
\right\rangle }{\left\langle \chi_{f}\right\vert U(t_{f},t_{i})\left\vert
\psi_{i}\right\rangle \left\vert s_{i}\right\rangle }\\
&  =\left[  \left\langle \psi_{I}\right\vert \left\langle +z\right\vert
\right]  A_{I}\left[  \left\vert \psi_{I}\right\rangle \left\vert
+z\right\rangle \right]  .\label{wvpI}%
\end{align}
For $A_{I}=\Pi_{I}$ (projector to the spatial wavefunction of arm $I,$
$\Pi_{I}\equiv\left\vert I\right\rangle \left\langle I\right\vert $) and
$A_{I}=\left(  \sigma_{x}\right)  _{I}$ (the spin component along the $x$
axis, where $I$ recalls the coupling takes place along path $I$), we obtain by
using Eq. (\ref{wvpI}):%
\begin{equation}
\Pi_{I}^{w}=1\qquad\left(  \sigma_{x}\right)  _{I}^{w}=0.\label{qccI}%
\end{equation}

Similarly, a quantum pointer coupled to $A_{II}$ can be placed along path
$II.\ $Its final quantum state $\left\vert \varphi_{II}(t_{f})\right\rangle $
is obtained after postselection (to the same state $\left\vert \chi
_{f}\right\rangle $) exactly as above, yielding%
\begin{equation}
\left\vert \varphi_{II}(t_{f})\right\rangle =\frac{1}{2}\exp\left(
-igA_{II}^{w}P_{II}\right)  \left\vert \varphi_{II}(t_{i})\right\rangle
\end{equation}
with%
\begin{equation}
A_{II}^{w}=\left[  \left\langle \psi_{II}\right\vert \left\langle
-z\right\vert \right]  A_{II}\left[  \left\vert \psi_{II}\right\rangle
\left\vert +z\right\rangle \right]  .
\end{equation}
This gives us for $A_{II}=\Pi_{II}$ and $A_{II}=\left(  \sigma_{x}\right)
_{II}$ respectively%
\begin{equation}
\Pi_{II}^{w}=0\qquad\left(  \sigma_{x}\right)  _{II}^{w}=1.\label{qccII}%
\end{equation}

The conjunction of Eqs. (\ref{qccI}) and (\ref{qccII}) defines the quantum
Cheshire Cat effect. Indeed, the quantum pointer coupled to the system on path
$I$ only detects the spatial wavefunction, but the interaction with the spin
component $\sigma_{x}$ has no effect.\ Conversely along arm $II$, a quantum
pointer picks up a shift due to the coupling with the system's spin component
$\sigma_{x}$ but the coupling to the spatial wavefunction along path $II$ has
no effect on the pointer.

Following our discussion on null weak values in Sec. \ref{nullwv}, the fact
that the particle cannot be seen on path $II$ means that the slight change
brought to the spatial wavefunction of the system by the interaction with the
weakly coupled pointer on path $II$ cannot be postselected in the state
$\left\vert \chi_{f}\right\rangle $ (the transition element $\left\langle
\chi_{f}\right\vert \Pi_{II}\left\vert \psi_{i}\right\rangle $ vanishes).
Similarly the spin component $\sigma_{x}$ is not detected on arm $I$ because
the postselected spin state $\left\vert +z\right\rangle $ cannot be reached by
the part $\sigma_{x}\left\vert +z\right\rangle $ of the spin state that has
been modified by coupling (the transition element $\left\langle
+z\right\vert \sigma_{x}\left\vert +z\right\rangle $ vanishes). Note that in
principle different weak measurements can be made jointly, on both arms, or
subsequently, on the same arm (since for an asymptotically weak interaction,
to lowest order, only the unperturbed part of the system state is taken into
account when the system interacts with a subsequent weak pointer).



\section{Experimental implementations of the Quantum Cheshire Cat}

\subsection{Experiment in a neutron interferometer \label{neutron}}

A few months after the publication of the Quantum Cheshire cat paper by
Aharonov et al \cite{aharonov-qcc}, an experimental realization of the QCC
effect was implemented \cite{A2014NatComm} with single neutrons in a
Mach-Zehnder-like triple-Laue interferometer. Due to strong experimental
constraints, the scheme employed in the experiment was significantly different
from the ideal QCC scheme described in Sec. \ref{qccT}.\ In particular, it was
not experimentally feasible to couple quantum pointers to the neutron inside
the interferometer; instead, the weak values were inferred from the
intensities of the detected neutron signal after postselection. While the weak
values determined experimentally were in excellent agreement with Eqs.
(\ref{qccI}) and (\ref{qccII}) defining the QCC effect, the fact that weak
measurements were not made (the weak values were instead inferred from
intensities measured after making transformations) has implications concerning
the validity of the observation of the QCC effect, as we now detail below.

Rather than summarizing the experiment as described in
Ref.\ \cite{A2014NatComm}, we will highlight the differences with the
theoretical QCC proposal. For simplicity we will employ the same pre and
postselected states employed in the theoretical proposal \cite{aharonov-qcc}
and in Sec.\ \ref{qccT} (in the experiment pre and postselected states were
inverted relative to the theoretical proposal, but this has no consequence for
our present discussion).

The crucial difference between the QCC theoretical proposal and the neutron
experiment is the lack of quantum pointers interacting with the neutron (or
the lack of additional degrees of freedom of the neutron that would act as
such pointers). Hence there is no such thing as the state vector $\left\vert
\varphi_{I}(t_{i})\right\rangle $ in equation (\ref{ini-qcc}) or in any of the
equations below equation (\ref{ini-qcc}). This is of course a radical
departure from the weak measurement formalism introduced in Sec. \ref{wv-f}.
Instead of relying on weak measurements, the neutron QCC experiment is based
on introducing external potentials along the arms. These interactions modify
the postselected intensity (ie, the neutrons detected in the postselected
state) relative to the intensity obtained without the interaction in that arm.

The spatial projector weak values are obtained by inserting an absorber on
arms $I$ or $II$. The absorber is not a quantum pointer: it is modeled by an
external decay potential $V_{j}=e^{-iM_{j}}$ where $j$ stands for either arm
$I$ or $II$; $M$ is the absorption coefficient. For $M$ small, some simple
manipulations lead to \cite{A2014NatComm}%
\begin{equation}
\mathcal{I}_{j}^{abs}=\left\vert \left\langle \chi_{f}\right\vert
U(t,t_{i})\left\vert \psi_{i}\right\rangle \left\vert +z\right\rangle
\right\vert ^{2}\left(  1-2M_{j}\Pi_{j}^{w}\right) =\mathcal{I}_{0}\left(
1-2M_{j}\Pi_{j}^{w}\right) \label{intensity-pres}%
\end{equation}
where $\mathcal{I}_{0}=\left\vert \left\langle \chi_{f}\right\vert
U(t,t_{i})\left\vert \psi_{i}\right\rangle \left\vert +z\right\rangle
\right\vert ^{2},$ obtained from Eqs.\ (\ref{incom}) and (\ref{qcc-ps}),
defines the detected reference intensity after postselection. Hence the
experimental observation of $\mathcal{I}_{j}^{abs}$ when the absorber is
placed on arm $j$ allows to extract the weak value $\Pi_{j}^{w}$.\ But no weak
measurement has been made: instead of a weak interaction, we have imposed a
strong interaction that happens with a small probability; instead of a pointer
whose state would reflect the effect of the coupled observable on the quantum
state of the pointer, we are inferring the presence of neutrons along path $I$
by the fact that the relative number of postselected neutrons decreases. Hence
a postselected neutron does not carry any signature of the interaction
(precisely because the neutrons that have interacted with the absorber on path
$I$ have been absorbed and have thus vanished). This does not entail that one
cannot conclude from Eq. (\ref{intensity-pres}) that the neutrons can only
reach postselection by going through path $I:$ indeed when an absorber is
placed along path $I$ this is reflected by the fact that $\mathcal{I}%
_{I}^{abs}/\mathcal{I}_{0}<1$ while an absorber along path $II$ has no effect
on the intensity, $\mathcal{I}_{2}^{abs}/\mathcal{I}_{0}=1.$ However this
conclusion does not rely on weak measurements nor on weak values but only on
the relative intensities, consistent with the fact that no coupling to a
quantum pointer has taken place: no observable has been weakly measured.
Instead, the conclusion $\Pi_{II}^{w}=0$ is made from the lack of backaction
of the strong interaction with the absorber on the postselected neutron
intensity (see Fig. 2).

For the spin component weak value the same objection can be made, now with
more serious consequences. The external potential employed is the one for a
magnetic moment in a magnetic field $B_{j}$ oriented along the $x$ axis,
$U_{j}=-\gamma\left(  \sigma_{x}\right)  _{j}B_{j}/2$ where $\gamma$ is the
gyromagnetic ratio and the index $j$ means that $B_{j}$ is non-zero only in a
region along arm $j$ (and thus affects only the magnetic moment $\gamma\left(
\sigma_{x}\right)  _{j}$ on arm $j$). Since $\int dtU_{j}=\alpha\left(
\sigma_{x}\right)  _{j}/2,$ where $\alpha$ is the precession angle induced by
the magnetic field, we have for small $\alpha$%
\begin{equation}
e^{-i\int dtU_{j}/\hbar}U(t_{w},t_{i})\left\vert \psi_{i}\right\rangle
\left\vert s_{i}\right\rangle =\left[  1+i\frac{\alpha}{2}\left(  \sigma
_{x}\right)  _{j}-\frac{\alpha^{2}}{8}\Pi_{j}\right]  \left(  \left\vert
\psi_{I}\right\rangle \left\vert s_{i}\right\rangle +\left\vert \psi
_{II}\right\rangle \left\vert s_{i}\right\rangle \right)  /\sqrt
{2}.\label{y10}%
\end{equation}
Note that the second order term is needed because postselection leads here to
intensities, not to pointer shifts [compare with Eq. (\ref{comp})].
Postselection indeed yields%
\begin{equation}
\mathcal{I}_{j}^{mag}=\left\vert \left\langle \chi_{f}\right\vert
U(t,t_{i})\left\vert \psi_{i}\right\rangle \left\vert +z\right\rangle
\right\vert ^{2}\left(  1+\frac{\alpha^{2}}{4}\left\vert \left(  \sigma
_{x}\right)  _{j}^{w}\right\vert ^{2}-\frac{\alpha^{2}}{4}\Pi_{j}^{w}\right)
\label{y11}%
\end{equation}
where $\mathcal{I}_{0}=\left\vert \left\langle \chi_{f}\right\vert
U(t,t_{i})\left\vert \psi_{i}\right\rangle \left\vert +z\right\rangle
\right\vert ^{2}$ is again the reference intensity and $\mathcal{I}_{j}^{mag}
$ the detected intensity when a magnetic field is applied on arm $j$.

In the neutron experiment, Eq. (\ref{y11}) is employed to compute $\left\vert
\left(  \sigma_{x}\right)  _{j}^{w}\right\vert $ from the observed relative
intensities $\mathcal{I}_{j}^{mag}/\mathcal{I}_{0}$. The term $\Pi_{j}^{w}$
was taken from the value inferred experimentally from Eq.
(\ref{intensity-pres}), though it would also have been consistent to use
$\delta_{I,j}$ instead.\ Indeed $\left(  \sigma_{x}\right)  _{j}^{2}=1_{j}$
(the identity along arm $j)$ but upon postselection $1_{j}\left(  \left\vert
\psi_{I}\right\rangle \left\vert s_{i}\right\rangle +\left\vert \psi
_{II}\right\rangle \left\vert s_{i}\right\rangle \right)  $ vanishes for $j=II
$. The upshot is that while the theoretical prediction $\left(  \sigma
_{x}\right)  _{I}^{w}=0$ can be recovered experimentally from $\mathcal{I}%
_{I}^{mag}$ by fitting Eq. (\ref{y11}), the magnetic field along arm $I$
nevertheless has an effect on the spin$\ \sigma_{x}$ through the last term
$-\alpha^{2}/4$ of Eq. (\ref{y11}), a point that was made in Ref.
\cite{stuckey2016}. In this particular experiment this effect can be claimed
to be systematic (in the sense that it doesn't depend on the field orientation
-- ie, which spin component couples with the field), but it is there
nonetheless: the appearance of higher order terms is generic when inferring
weak values from intensity measurements, rather than making genuine weak measurements.

\begin{figure}[tb]
\includegraphics[height=8cm]{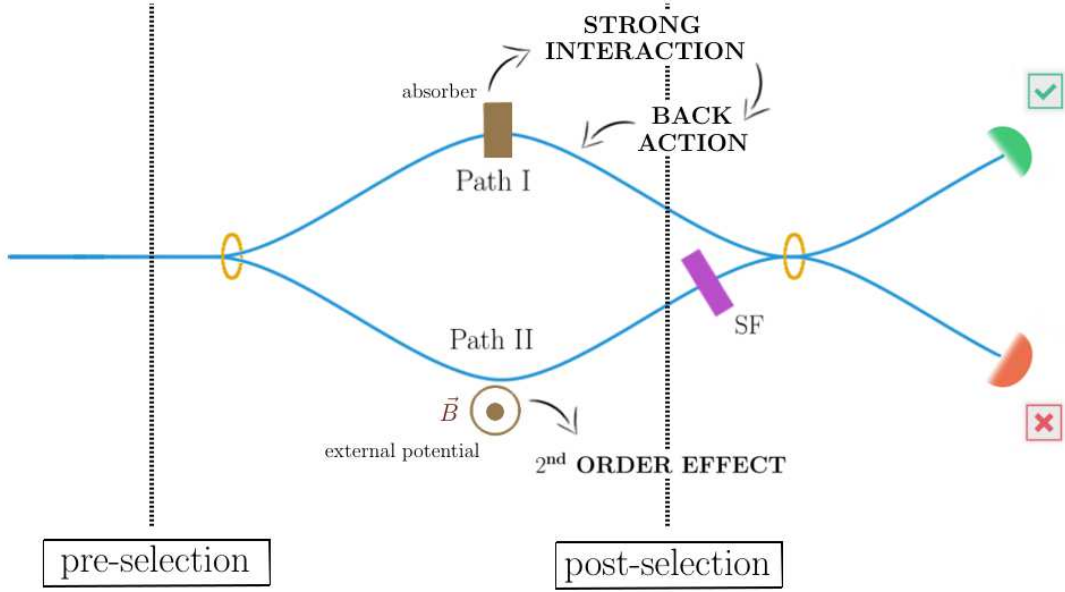}\caption{Schematic representation of the
differences between the ideal Quantum Cheshire Cat setup proposed
theoretically and the experimental implementation with neutrons
\cite{A2014NatComm}. These differences also apply to a similar experiment
performed with photons in a Sagnac-like interferometer \cite{QCC-photon} (see
text).}%
\label{fig2}%
\end{figure}

\subsection{Experiment with single photons \label{photons}}

A QCC experiment with photons was carried out very recently \cite{QCC-photon}.
The authors make clear from the start that their setup is based on the one
implemented in the neutron experiment, where the photon polarization takes the
place of the neutron spin (the polarization state along the interferometer
arms is $\left\vert H\right\rangle $ or $\left\vert V\right\rangle $ instead
of the spin states $\left\vert +z\right\rangle $ and $\left\vert
-z\right\rangle $, and the circular polarisation weak value replaces $\left(
\sigma_{x}\right)  _{j}^{w})$. Indeed, leaving aside the specificities of the
photon experiment (use of a Sagnac interferometer, coincidence detection with
the heralded photon), the scheme employed in Ref. \cite{QCC-photon} in order
to display the QCC effect is identical to the neutron experiment: the presence
weak value $\Pi_{j}^{w}$ is inferred from the photon counts with/without an
absorber placed along the arms, and the polarization weak value is inferred
from photon counts with/without a rotation of the polarization induced by half
wave plates. This similarity can be seen directly by comparing Eqs. (3) and
(6) of Ref. \cite{QCC-photon} with Eqs. (\ref{intensity-pres}) and (\ref{y11})
given above for the neutron experiment respectively.

Hence, our remarks made above for the neutron experiment also hold for this
photon experiment: no weak measurements were made, and inferring weak values
from measured intensities does not give an unambiguous demonstration of the
Cheshire cat effect. It is noteworthy that the authors of Ref.
\cite{QCC-photon} point out in their Conclusion that their observed
measurement statistics can be understood without recourse to weak values.

\section{Criticism of the QCC in the literature}

\subsection{General remarks}

The Quantum Cheshire Cat scheme has been criticized by several
groups \cite{CCNJP,deraedt2015,atherton2015,stuckey2016,soko2016}.\ At the
basis of the criticism there is the underlying idea that claiming
separation of a property from a particle is preposterous. However
the criticism was not always carefully formulated. Most of the works did
not clearly discriminate the original theoretical proposal from the
neutron experimental implementation (which as we have shown above
are very different). They were also not rigorous when making
assertions concerning the disembodiment, often using terms in plain
English (which can be at best an interpretation of the QCC\ effect)
rather then scrutinizing the technical definition of the QCC effect
in terms of weak measurements; actually most of these works
with the exception of Ref. \cite{soko2016} did not base their
arguments on the characteristics of the weak measurements framework,
that was generally ignored.

As a result, the criticism did not deliver clarification. We study below
the main arguments that were given in the criticisms and assess
their relevance. The first class of arguments involves the detected
intensities, a second type of arguments invokes interference, while
other arguments attempt to discard the possibility of disembodiment.

\subsection{Intensities}

The work by Stuckey et al \cite{stuckey2016} criticizes the QCC neutron
experiment only, on the basis that the detected intensities when the spin weak
value is $\left(  \sigma_{x}\right)  _{I}^{w}=0$ still shows the effect of the
magnetic field (they do accept that the absorber's effect on the intensities
is a proof that the particle did not take path $II$). Their agument starts
from the identity%
\begin{equation}
e^{-i\alpha\sigma_{x}/2}=\cos\frac{\alpha}{2}I- i \sin\frac{\alpha}{2}\sigma_{x}.
\end{equation}
Stuckey et al point out that the magnetic field has an effect on the
intensities through the $\cos\frac{\alpha}{2}$ term: to lowest order this term
is quadratic, exactly as the term containing the weak value contribution. So
they conclude that this contradicts the statement according to which the spin
component $\sigma_{x}$ did not travel through arm $I$. This argument is
technically correct as far as intensities are concerned; it is equivalent to
the observation we made below Eq. (\ref{y11}). However this argument does not
involve weak measurements (although we know from our discussion above that the
QCC is defined from weak measurements that leave the post-selected intensities
of the system undisturbed), nor does Ref. \cite{stuckey2016} propose an
alternative framework for a non-destructive measurement that would ascribe a
non-zero value to the spin component on arm $I$, despite the fact that
$\left(  \sigma_{x}\right)  _{I}^{w}=0$. Still it is valuable to remark that
Stuckey et al \cite{stuckey2016} leave open the possibility of observing a
genuine QCC.

In a very different work Atherton et al \cite{atherton2015}
performed a classical optics experiment based on a Mach-Zehnder type
interferometer. They start with polarized beams and compare the
intensities obtained on one of the output ports with or without an
absorber or a polarization rotating plate inserted on either arms
$I$ or $II$ (a polarizer ensures postselection of the detected
beam).\ This setup indeed mimicks the neutron experiment described
in Sec. \ref{neutron}. The observed electric field intensities
display the same behavior as the neutron counts. On this basis,
Atherton et al conclude that there is nothing quantum about the QCC
effect, that in their view is an \textquotedblleft
illusion\textquotedblright. This classical physics experiment has
the merit of highlighting the fact that the observed intensities in
the neutron experiment do not need to be interpreted with the weak
measurements formalism, since no quantum weak values can be obtained
with classical beams. However attempting to draw conclusions on the
value of quantum observables from a classical optics experiment is
an impossible task: the classical beams travel through both arms, so
what could the Cheshire cat effect mean in this context? Instead a
single neutron obeys the standard quantum rule of projective
measurements and cannot be found simultaneously on both arms.
Atherton et. al. do not define precisely the analogue of the quantum
observable $\sigma_x$ and there is no mention of anything that can
play the role of weak values in their work, although they are
necessary to precisely define the QCC effect. Therefore the findings
of Ref. \cite{atherton2015} are only relevant to the specific
implementation of the neutron experiment summarized in Sec.\
\ref{neutron}, not to the QCC\ effect itself.

\subsection{Interference}

In an interesting paper, Correa et al \cite{CCNJP} assert that the QCC effect
arises from \textquotedblleft simple quantum interference\textquotedblright.
They mean by that statement that a quantum pointer remaining unchanged (as
this happens for vanishing weak values) results from the interference of
almost perfectly overlapping final pointer states. This remark is
uncontroversial -- this is of course the way weak measurements work in general
[see Eq. (\ref{finalps})], as shown early in a 1989 work \cite{duck89}.
Significantly, Correa et al. explicitly introduce the quantum pointer states
(our states $\left\vert \varphi(t_{i})\right\rangle $ of Secs. \ref{wv-f} and
\ref{qccT}) but treat the pointer-system interactions in an ad-hoc way, by
stating in words how these pointer states are transformed, rather than
introducing an interaction Hamiltonian such as our equation (\ref{Hint}).
Hence weak values do not explicitly appear in their treatment, but the motion
of the weakly coupled pointer states are recovered by the superposition of the
quantum state after postselection, see our Eq. (\ref{finalpspos}) above.

However in our view the Quantum Cheshire Cat\ effect does not depend on the
underlying mathematical formulation of the pointers motion, but on giving a
physical meaning to these motions, as we now discuss for the most salient
physical property claimed to characterize the QCC effect, disembodiment.

\subsection{Disembodiment}

The main objective of the criticisms seems to undermine the claim
made in the original proposal by Aharonov et al.\ that Eqs.
(\ref{qccI})-(\ref{qccII}) could imply disembodiment of the particle
from one of its properties. Correa et al \cite{CCNJP} write that
their interference argument allows them to interpret the QCC
phenomenon without appealing to disembodiment, but they do not
produce a full reasoning that would support this claim. Indeed, the
interference account of the quantum pointer dynamics does not
endorse nor disprove the \textquotedblleft
disembodiment\textquotedblright\ claim: the superposition of state
vectors is a Hilbert space feature, that at least according to
standard quantum mechanics is only a mathematical description aimed
to compute probabilities. The authors of Ref. \cite{CCNJP} do not
elaborate on whether their argument implies going beyond this
standard view, for example by endowing the state vectors with some
ontological features that would then propagate along both arms.
Instead, whether the motion of the quantum pointers weakly coupled
to the particle's position or spin component is indicative of
disembodiment depends on a framework ascribing properties to a
quantum system in the absence of a projective measurement. The weak
measurements formalism constitutes such a framework, and discarding
the possibility of disembodiment implies either replacing the WM
formalism with some alternative proposition, or refuse that quantum
properties can be defined beyond projective measurements. This point
will be further discussed in Sec. \ref{discuss}.

Michielsen et al \cite{deraedt2015} also objected on disembodiment by running
numerical experiments.\ Strictly speaking their work only applies to the
neutron experiment. They simulate the observed interferences with/without
absorbers/spin rotators on arms $I$ and $II$.\ The simulation is based on a
discrete event learning model, in which particles act as messengers in such a
way that the particle counts in the outgoing ports of a Mach-Zehnder
interferometer quickly converge towards the quantum probabilities. Such a
model was previously employed to reproduce intensities in neutron
interferometric experiments \cite{deraedt2012}, so Ref \cite{deraedt2015} is
an extension of that previous work so as to include the absorber/rotator
interactions. In any case Michielsen et al. do not consider quantum pointers
and weak measurements in their model, and while it would be interesting to
investigate if the Quantum Cheshire Cat can be properly formulated within the
discrete event learning model by including the coupled quantum pointers
explicitly, their results are not relevant to the QCC effect as we have
properly defined it.

\section{Discussion\label{discuss}}

\subsection{General remarks}

The main property characterizing the Quantum Cheshire Cat seen here
is that for a fixed initial and final state of the system (photon,
neutron), the state of a quantum pointer weakly coupled to the
system's spatial wavefunction is modified for a pointer placed along
path $I$ but not for a pointer placed along path $II$.\ If the
quantum pointer is coupled to the spin component $\sigma_{x}$
instead, the opposite behavior is obtained. Since neither the
neutron nor the single photon experiments have recourse to weak
pointers, criticizing the experiments done so far as not having
realized the QCC is legitimate.\ However we have seen that some
authors of the criticism did not clearly discriminate the
experimental implementations from the ideal QCC.\ For instance
Atherton et al \cite{atherton2015} and Michielsen et al
\cite{deraedt2015} criticize the neutron experiment, but from there
cast suspicion on the ideal theoretical scheme as being an
\textquotedblleft illusion\textquotedblright.

While doing so is strictly speaking inconsistent, the common element
underlying the criticism of the Quantum Chehsire Cat formulated in
Refs \cite{CCNJP,deraedt2015,atherton2015,stuckey2016} is to refute
the idea of a spatial separation between the particle and one of its
properties. Unfortunately, rather than starting from the technical
definition of the QCC and from there prove that this definition
cannot imply a spatial separation, Refs
\cite{CCNJP,deraedt2015,atherton2015,stuckey2016} do not specify the
 assumptions they make concerning the possibility of ascribing  properties to
a quantum system in the absence of a projective measurement. The
weak measurements formalism proposes a framework accounting for such
properties, and from there a technical definition for spatial
separation is obtained in terms of weak values. This is arguably different than
taking the term ``spatial separation'' in a literal sense, that would disregard
the well-known conceptual difficulties of the standard quantum
formalism in giving an unambiguous account of the physical state and
properties of a system. These difficulties are ultimately due to the
fact that contrary to classical mechanics or classical optics, the
relationship between the theoretical terms of quantum mechanics and
physical reality are unknown, and most often denied. Hence relying
on pointers to assess the value of the property of a quantum system
is crucial; this in turn hinges on employing an explicit conceptual
and interpretative framework.

\subsection{Technical and conceptual aspects}

The Quantum Cheshire Cat is technically defined by Eqs. (\ref{qccI})
and (\ref{qccII}) in the context of weak measurements with
postselection that do not affect the coherence of the system. So the
first question is whether one can have Eqs. (\ref{qccI}) and
(\ref{qccII}) (in the context of weak measurements) while still
being able to assess the particle can be found on path $II$, or that
spin component $\sigma_{x}$ can be found along path $I$ (conditioned
on successful postselection). In order to answer this question, the
authors of Refs. \cite{CCNJP,deraedt2015,atherton2015,stuckey2016}
do not take into account the fact that some type of measurement or
interaction (presumably different from the weak measurement
protocol) needs to be proposed. Otherwise it is impossible to assert
anything about the system properties. In particular, relying on the
state vector, as done in \cite{CCNJP,stuckey2016}, is insufficient:
every quantum physicist agrees that the total state vector is in a
superposition state along both paths, and that postselection will
imply a certain correlation due to interference. But the quantum
axiomatics remain silent on the meaning of the state vector, that
may be taken as a simple computational tool (this is the standard
view conveyed in textbooks), or as an element of a more elaborate
ontology, but clearly not in a literal manner as representing a
classical field that would propagate simultaneously along both arms
of the interferometer.

\subsection{Interpretations}

This leads us to the second question: does the technical definition
of the QCC recalled in the preceding paragraph necessarily imply
some form of spatial separation, or disembodiment? The answer
depends on how the quantum pointer's motions are interpreted.\
Indeed, the pointers measure weak values, and null weak values  are
null transition elements.\ And transition elements are
well-defined quantities in standard quantum mechanics.

Now having pointers measuring transition elements does not fit with the
eigenstate-eigenvalue link (by which the value of the property represented by
the observable is associated with a quantum state of the system). This would
be impossible given the aim of the WM scheme, as recalled at the beginning of
Sec. \ref{wv-f}: a given weak value is not associated with a given state of
the system, but with the transition of the (time-evolved) preselected state to
the postselected state induced by the weak coupling between system and pointer
observables. As we have seen in Sec. \ref{nullwv}, vanishing transition
amplitudes imply that the final postselected state cannot be reached by the
part of the system state that has been perturbed by the weak interaction with
the quantum pointer.

Hence $\Pi_{II}^{w}=0$ implies that the transformation generated on the
preselected state by premeasuring $\Pi_{II}$ does not reach the postselected
state.\ If one upholds an interpretation in which a property relies on projection
to an eigenstate, then measuring a vanishing transition element
has no bearing on a statement concerning property ascription.\ This is the
criticism developed by Sokolovski \cite{soko2016}. In the terminology
of Ref. \cite{soko2016}, the transition 
amplitudes belong to ``virtual paths'', that do not describe the real path of a particle. Indeed,
according to this view, only a strong projective measurement can tell us if
the particle is or not in arm $II$. A projective measurement creates a ``real path'' that precludes the possibility of
measuring any other property on the arm in which the particle is not found (otherwise the uncertainty principle would be violated), so that by that account
a real path cannot accommodate the idea of spatial separation. 
If we do not perform a projective measurement, but
measure instead a null weak value, then we know that the transition element
$\left\langle \chi_{f}\right\vert \Pi_{II}\left\vert \psi_{i}\right\rangle $
vanishes, but this only characterizes the reaction of the system to a small perturbation
and should not be taken as a measurement of the position of the
particle.

While this point of view is consistent, it is also possible to go further in
interpreting transition elements as characterizing system properties. The null weak value operationally means that the pointer state corresponding to the particles detected in the postselected state is unaffected by the presence of weak interactions. Accordingly $\Pi_{II}^{w}=0$ implies that the
particle's presence cannot be detected by a quantum pointer weakly interacting
with the spatial wavefunction on path $II$ \emph{and }be detected in the
chosen postselection state $\left\vert \chi_{f}\right\rangle $ [Eq.
(\ref{qcc-ps})]: such a correlation is forbidden by standard quantum
mechanics, as the transition element $\left\langle \chi_{f}\right\vert
\Pi_{II}\left\vert \psi_{i}\right\rangle $ vanishes. On this basis it is
possible to uphold that the final state cannot have been reached by taking path
$II$ (otherwise the weakly coupled pointer along arm $II$ would have been
displaced upon postselection).\ In a crude sense, we can say that the particle
has not been in arm $II$.\

The same reasoning can be made concerning the spin component. $\left(
\sigma_{x}\right)  _{I}^{w}=0$ because the transition element $\left\langle
+z\right\vert \left(  \sigma_{x}\right)  _{I}\left\vert +z \right\rangle $
vanishes. A vanishing transition element means that the transformed part of
the spin state resulting from coupling $\sigma_{x}$ weakly to a quantum
pointer on path $I$ is orthogonal to the postselected state (hence the
postselected cannot be reached by the fraction of the system state coupled to
the pointer). Therefore $\sigma_{x}$ cannot be found on path $I$ in this
sense: the premeasurement of $\sigma_{x}$ on path $I$ slightly changes the
spin state, but for the stipulated postselection, this slight change has no
effect, leaving the state of the quantum pointer undisturbed. Note that this
is very different from the projective measurement process yielding a null
eigenvalue, given that a vanishing eigenvalue is associated with a particular
eigenstate (see Sec. \ref{nullwv}).

 It should be
stressed that relying on transition elements to assess the value of properties in pre/post-selected systems does not necessarily call for paradoxes. This can be understood as the
confirmation that the effect of the superposition principle (or sum over
paths) can be observed by a local weak coupling of a system observable with a
quantum pointer.\ In general, the transition element on either path will be
not vanishing, yielding an observable effect on the pointers placed on both
arms of the interferometer. However for specific choices of pre/postselected
states, a given system observable may generate a transition to the final state
only along a given arm, while such a transition cannot take place along the other.

\section{Conclusion}

In this paper, we have bridged the gap, both at a theoretical and at a
conceptual level, between the original Quantum Cheshire Cat proposal
\cite{aharonov-qcc}, and the various experimental realizations and criticisms
that have been formulated. In doing so, we have clarified the meaning of the
QCC effect and dispelled the considerable degree of confusion that was seen to
arise from the raised criticisms of the Quantum Cheshire Cat proposal.

We have argued that ``disembodiment'' can be said to hold if it is defined in
terms of transition elements for the system observables.  As we have seen, a quantum pointer after postselection detects the
system observable to which it is weakly coupled only when the relevant
transition element for that observable does not vanish. By a suitable choice of pre and post-selected states, the spatial
wavefunction can only be detected by a quantum pointer placed on one of the paths (and not the other) while the spin component is only seen on the other path.

We conclude by summarizing our main results:

\begin{itemize}
\item The Quantum Cheshire Cat effect is a well defined quantum feature
derived from the standard quantum formalism for pre- and
post-selected states of a system; the interpretation of the effect in terms of spatial 
separation of a particle from one of its properties hinges on the issue of the relation between property ascription and weakly coupled pointers;

\item The QCC effect as predicted theoretically has not yet been
experimentally observed, as the experimental realizations done so far have not
been able to properly implement the weak measurement protocol;

\item Most of the works criticizing the QCC effect did not introduce a proper
framework in order to analyze the issue of spatial separation of a
quantum particle from one of its properties in a pre and
postselected situation, so their criticism is incomplete.
\end{itemize}

\vspace{1cm}

\textbf{Acknowledgements} Partial support from the Templeton
Foundation (Project 57758) is gratefully acknowledged. AM thanks the Institute
for Quantum Studies (Chapman University) for hospitality at the time this work was completed.

\vspace{1cm}




\end{document}